\documentclass[12pt,letterpaper]{article}

\usepackage[margin=1in]{geometry}
\usepackage{amsmath,amssymb}
\usepackage{booktabs}
\usepackage{multirow}
\usepackage[round,authoryear]{natbib}
\usepackage{indentfirst}
\usepackage{placeins}
\usepackage{setspace}
\usepackage{hyperref}
\usepackage{microtype}
\usepackage{parskip}
\usepackage{graphicx}
\usepackage{float}

\hypersetup{
    colorlinks=true,
    linkcolor=blue,
    citecolor=blue,
    urlcolor=blue,
    pdftitle={MACD},
    pdfauthor={Luyun Lin et al.}
}

\begin{document}

\begin{titlepage}
\centering
\vspace*{1.5cm}

{\Large\bfseries A Volume-Price-Adjusted MACD Trading Strategy with Sensitivity Calibration for U.S. Equity Indices}

\vspace{1.5cm}



\begin{center}

{\normalsize
\begin{minipage}[t]{0.30\textwidth}
\centering
Luyun Lin\\[4pt]
\textit{Independent Researcher}\\[4pt]
Dallas, TX, USA
\end{minipage}
\hfill
\begin{minipage}[t]{0.30\textwidth}
\centering
Lixing Lin\\[4pt]
\textit{Yale University}\\[4pt]
New Haven, CT, USA
\end{minipage}
\hfill
\begin{minipage}[t]{0.30\textwidth}
\centering
Zhen Zhang\\[4pt]
\textit{Independent Researcher}\\[4pt]
Jersey City, NJ, USA
\end{minipage}
}

\vspace{0.5cm}

{\normalsize
\begin{minipage}[t]{0.40\textwidth}
\centering
Moxuan Zheng\\[4pt]
\textit{Independent Researcher}\\[4pt]
Jersey City, NJ, USA
\end{minipage}
\hspace{0.08\textwidth}
\begin{minipage}[t]{0.40\textwidth}
\centering
Yiqing Wang\textsuperscript{*}\\[4pt]
\textit{Independent Researcher}\\[4pt]
Dallas, TX, USA
\end{minipage}
}

\end{center}


\renewcommand{\thefootnote}{\fnsymbol{footnote}}
\setcounter{footnote}{0}

\footnotetext[1]{Corresponding author. Contact Email: 
\href{mailto:woshilucy712@gmail.com}{woshilucy712@gmail.com}.}
\vspace{1cm}

\begin{abstract}
Traditional moving average convergence divergence (MACD) trading rules are often constrained by signal lag and susceptibility to false signals. To address these limitations, this study develops a volume-price-adjusted MACD (VP-MACD) framework that incorporates volume, volatility, and intraday price structure into the conventional indicator, and introduces a sensitivity parameter, $\lambda$, to allow earlier trade entry and improve responsiveness to market movements. Using the S\&P 500, Nasdaq-100, and Dow Jones Industrial Average as representative U.S. equity indices, the model is calibrated over historical records from 2018 to 2022 and evaluated out of sample over 2023 to February 2026. The results indicate that the proposed framework generally delivers better economic performance than the baseline MACD strategy in terms of profitability, risk-adjusted return, and downside-risk control, while generating fewer but more selective trading signals. These findings suggest that incorporating additional market information into technical trading rules may enhance signal quality in U.S. equity index markets.
\end{abstract}



\vspace{0.5cm}

\noindent
\begin{minipage}{0.95\textwidth}
\end{minipage}

\vspace{0.5cm}

\end{titlepage}


\section{Introduction}
\label{sec:intro}

Technical analysis has long been used by both market practitioners and academic researchers as a framework for generating trading signals directly from market data, without relying on firm-level fundamental information \citep{edwards2018technical}. Among the indicators developed within this tradition, the Moving Average Convergence Divergence (MACD) remains one of the most widely used because of its simplicity, interpretability, and applicability across asset classes and time horizons \citep{murphy1999technical}. Despite these advantages, the conventional MACD suffers from two well-recognized limitations. First, because it is constructed solely from smoothed exponential moving averages of closing prices, its signals respond to market movements with an inherent lag, often delaying entry and exit decisions relative to actual trend reversals. Second, the indicator does not incorporate trading volume or intraday price information, making it more susceptible to false crossover signals when observed price changes are not supported by sufficient market participation or directional strength.

Prior studies have attempted to address these limitations either by optimizing MACD parameters for specific market \citep{kang2021improving,agudelo2020macd} or by augmenting the signal with additional technical indicators \citep{wang2018predicting,chio2022comparative}. However, parameter optimization is highly sensitive to the choice of calibration window and may lead a risk of in-sample overfitting. Indicator augmentation can improve signal quality, but it often results in a fragmented analytical structure rather than a coherent and unified framework.

This study addresses these gaps by proposing a volume-price-adjusted MACD (VP-MACD) framework embeding volume, volatility, and price-structure information directly into the indicator construction remains comparatively underexplored. The core modification replaces the conventional  closing-price input with an adjusted price series that jointly weights volume intensity, intraday volatility, and candlestick body ratio, thereby 
enriching the informational basis of the indicator without altering itsfundamental architecture. In addition, a sensitivity parameter, $\lambda$, 
is introduced to relax the crossover condition, enabling earlier entry signals and improving responsiveness to nascent trend changes. The 
framework is calibrated on S\&P~500, Nasdaq-100, and Dow Jones Industrial Average data from 2018 to 2022 and evaluated out-of-sample over 2023 to 
February 2026.

The main contributions of this study are threefold. First, a unified structural enhancement of MACD is developed that integrates volume, 
volatility, and intraday price structure into a single adjusted price series, avoiding the fragmentation inherent in multi-indicator 
combinations. Second, a tunable sensitivity parameter $\lambda$ is introduced to systematically govern the trade-off between signal 
selectivity and entry timing. Third, the framework is evaluated across three major U.S.\ equity indices using a strict out-of-sample design, 
providing evidence on both performance improvements and their statistical 
significance.

The remainder of this paper is organized as follows. Section~\ref{sec:lit} 
reviews the relevant literature on MACD and technical trading rules. Section~\ref{sec:method} develops the VP-MACD framework and describes the back-testing design. Section~\ref{sec:results} presents and discusses the empirical results. Section~\ref{sec:conclusions} concludes and discusses the future direction.

\section{Literature Review}
\label{sec:lit}

\subsection{Market Efficiency and Technical Analysis}

The empirical viability of technical trading rules is closely connected to the debate on market efficiency. Under the weak-form Efficient Market 
Hypothesis (EMH), past price information is already reflected in current prices, leaving no systematic role for technical analysis 
\citep{fama1970efficient}. Behavioral finance offers a partial counter-argument, attributing recurring price patterns to investor sentiment and feedback dynamics \citep{shiller2003efficient}. The Adaptive Markets Hypothesis (AMH) 
reconciles these views by positing that the profitability of technical rules evolves over time and across market environments \citep{lo2004adaptive}. This time-varying perspective implies that indicators such as MACD may retain predictive value in specific markets or regimes even when efficiency broadly holds.

\subsection{Empirical Evidence on MACD Performance}

\citet{brock1992simple} provide an early benchmark demonstrating that simple 
technical trading rules can generate returns inconsistent with weak-form 
efficiency. A broader survey by \citet{park2004profitability} finds that technical 
rules yield economically significant profits in certain markets, particularly foreign exchange and futures, though results are sensitive to asset class, time period, and methodology. Evidence specific to MACD reveals comparable heterogeneity. Using a cross-country sample of 1,336 
companies across 75 countries, \citet{anghel2015stock} show that informational 
efficiency varies substantially, with the United States ranking only 59th, suggesting that even large and liquid equity markets may not fully impound historical price information. This finding is directly relevant to the present study.

Evidence from individual markets reinforces the conclusion that MACD performance is market-dependent rather than universally stable. 
\citet{chong2008technical} find that MACD-based strategies outperform passive benchmarks in the London market, whereas \citet{chong2014revisiting} document considerable variation across OECD markets, with no consistent outperformance in the United States or Japan. Emerging-market results 
are similarly mixed: \citet{chen2021effectiveness} reports positive outcomes for the Chinese equity market, while \citet{hejase2017technical} find no significant improvement in Lebanon. \citet{gunasekarage2001profitability} provide additional evidence that moving-average rules can be profitable in South Asian markets. Collectively, these findings suggest that MACD effectiveness is shaped by market-specific characteristics—including liquidity, investor composition, and regime conditions—rather than by a simple developed-versus-emerging distinction.

Beyond equities, MACD-based strategies have demonstrated adaptability 
across non-traditional asset classes. \citet{cohen2022complexity} documents robust 
predictive performance of optimized technical indicators in high-volatility cryptocurrency markets, and \citet{kang2026optimal} confirm 
the effectiveness of MACD-based rules in capturing trend reversals in gold markets, though performance remains sensitive to market structure 
and implementation design.

\subsection{Limitations of Traditional MACD and Enhancement}
\label{sec:lit_macd_lim}

Despite its broad application, the traditional MACD is constrained by two structural weaknesses. First, because the indicator is constructed 
entirely from exponential moving averages of historical closing prices, it reacts with an inherent delay relative to underlying market dynamics \citep{murphy1999technical}. \citet{wang2018predicting} attribute this to structural latency in the smoothing process, which limits the indicator's responsiveness to rapidly changing conditions. Second, the conventional 
specification ignores trading volume and intraday price structure. \citet{kang2021improving} documents that the standard (12, 26, 9) parameterization frequently generates false signals in choppy markets, and \citet{chen2021effectiveness} finds that MACD effectiveness deteriorates during consolidating or 
trend-reversal periods. These weaknesses help explain the heterogeneity in empirical findings documented above.

Prior research has pursued two main remediation strategies. The first relies on parameter optimization: \citet{kang2021improving} shows that 
market-specific parameter selection improves performance in the Nikkei 225 futures market, and \citet{agudelo2020macd} apply machine-learning methods to automate this process. However, market-specific calibration 
carries a well-recognized risk of in-sample overfitting and data snooping \citep{anghel2015stock}, and optimized parameters tend to have limited transferability across regimes. The second strategy augments 
MACD with supplementary market information. \citet{wang2018predicting} propose 
a volatility-adjusted MACD, and \citet{chio2022comparative} show that combining 
MACD with volume- and volatility-derived signals improves win rate and risk-adjusted performance in U.S.\ equities. Classical technical 
analysis has long recognized trading volume as a key confirmation variable for trend validity \citep{edwards2018technical}, a view supported by 
empirical evidence on the price-volume relationship in equity markets 
\citep{sanvicente2015price}. Nevertheless, these augmentation approaches typically combine separate indicators rather than integrating additional information into the indicator's core construction, resulting in fragmented rather than unified frameworks.

The present study addresses these limitations by embedding volume, intraday volatility, and candlestick price structure directly into the 
MACD price input, and by introducing a sensitivity parameter $\lambda$ to govern entry timing. The structural approach, volume-price adjusted MACD (VP-MACD),  improve signal robustness and practical transferability without
relying primarily on market-specific parameter optimization.

\section{Methodology}
\label{sec:method}

This section outlines the methodological framework. We
first introduce the MACD indicator and its underlying components,
followed by the specification of trading rules based on signal line
crossovers. We then examine the limitations of the traditional MACD,
including signal lag and susceptibility to false signals. Motivated by
these limitations, we propose a volume-price-adjusted MACD (VP-MACD)
model that integrates volume, price volatility, and structural price
information into the conventional framework. Finally, the backtesting
design and evaluation metrics are presented to assess the effectiveness
of the proposed approach.

\subsection{MACD Technical Indicator Definition}
\label{sec:macd_def}

The Moving Average (MA) is one of the most fundamental tools in technical analysis. The most basic variation is the Simple Moving Average (SMA), which is calculated by taking the arithmetic mean of an asset's closing prices over a specific number of periods. It is expressed as:
\begin{equation}
  \mathrm{SMA}_{t} = \frac{1}{n} \sum_{i=0}^{n-1} P_{t-i}
  \label{eq:sma}
\end{equation}
where $P_{t-i}$ denotes the closing price at time \(t-i\), and $n$ represents the lookback period, defined as the number of trading days in this study.

Moving averages are essentially trend-following devices designed to identify the beginning or reversal of a market trend by filtering out short-term price noise \citep{murphy1999technical}. In practice, investors frequently employ different benchmarks to gauge market sentiment: short-term trends are often captured by the 5-day MA (MA5), while medium and long-term sentiments are represented by the 20-day (MA20) and 60-day (MA60) averages, respectively. An upward-sloping MA, with price action sustained above the line, typically signifies a dominant buyer’s market. However, the SMA’s reliance on equal weighting for all data points introduces a significant drawback: it reacts slowly to sudden market shifts, a phenomenon commonly referred to as lag. This lag can result in delayed signals, which may be particularly problematic in volatile markets.

To partially mitigate this, the Exponential Moving Average (EMA) assigns greater weight to more recent price observations. This weighting scheme improves responsiveness to recent market movements while preserving the smoothing function of a moving average. Therefore, the EMA provides a useful foundation for constructing momentum indicators that compare price dynamics across different time scales, such as the MACD. The EMA is defined recursively as

\begin{equation}
  \mathrm{EMA}_{t} = \alpha\, P_{t} + (1 - \alpha)\,\mathrm{EMA}_{t-1}
  \label{eq:ema}
\end{equation}
where $\alpha$ is the smoothing factor, typically defined as 
$\alpha = 2/(n+1)$ and n is the lookback period of the exponential moving average, defined as the number of trading days in this study.

Although EMA improves the responsiveness of traditional moving averages, moving averages themselves primarily describe the direction of price trends and do not inherently measure trend strength, momentum acceleration, or potential exhaustion. Building on the EMA framework, the Moving Average Convergence Divergence (MACD) was originally developed by Gerald Appel in 1979 as a trend-following momentum indicator \citep{appel1979moving}. Since then, MACD has become a cornerstone of technical analysis. The indicator is constructed from the interaction between short-term and long-term exponential moving averages to capture price dynamics across different time scales. The calculation of MACD involves three adjustable time-period parameters, typically denoted as $(n_{fast}, n_{slow}, n_{signal})$. The standard configuration, commonly presented in technical analysis literature, is the (12, 26, 9) setting \citep{murphy1999technical}. This configuration uses a 12-day EMA for short-term momentum, a 26-day EMA for longer-term trend tracking, and a 9-day EMA of the MACD line itself as a smoothing signal. Since MACD relies solely on price time series, it can be applied across a wide range of asset classes, including equities, commodities, options, and cryptocurrencies, as well as across different time frequencies such as daily and intraday intervals.

The MACD is composed of three main components. 
\begin{itemize}
    \item The MACD line: defined as the difference between short-term and long-term EMAs, oscillates around zero. Its magnitude reflects the degree of separation between the two trends: values near zero suggest weakening momentum, whereas larger deviations from zero indicate stronger momentum
    \begin{equation}
    \mathrm{MACD}_{t} = \mathrm{EMA}_{12}(P_{t}) - \mathrm{EMA}_{26}(P_{t})
    \label{eq:macd}
    \end{equation}
    
    \item the signal line: defined as the exponential moving average of the MACD line, typically computed using a 9-period EMA
    \begin{equation}
      \mathrm{Signal}_{t} = \mathrm{EMA}_{9}(\mathrm{MACD}_{t})
      \label{eq:signal}
    \end{equation}

    \item Histogram: defined as the difference between the MACD line and the signal line, providing a measure of momentum strength and direction. When the histogram is above the zero line but moves toward it, this indicates weakening bullish momentum and a potential trend reversal may occur

    \begin{equation}
  \mathrm{Histogram}_{t} = \mathrm{MACD}_{t} - \mathrm{Signal}_{t}
  \label{eq:hist}
\end{equation}
\end{itemize}

\subsection{MACD Trading Rules}
\label{sec:trading_rules}
Based on the MACD indicator defined in Section 3.1, this section develops the corresponding trading rules. The rules are constructed from the interaction between the MACD line and the signal line, as well as their relationship with the zero line and price movements.

\vspace{\baselineskip}

\noindent\textbf{MACD signal line crossover strategy}


A common trading rule is based on the crossover between the MACD line
and the signal line. A buy signal is generated when the MACD line
crosses above the signal line, indicating increasing upward momentum.
Conversely, a sell signal is triggered when the MACD line crosses below
the signal line, suggesting weakening momentum. Trading signals are
defined as:
\begin{equation}
  \text{Signal} =
  \begin{cases}
    \text{Buy},  & \text{if } \mathrm{MACD}_{t-1} \leq \mathrm{Signal}_{t-1}
                   \text{ and } \mathrm{MACD}_{t} > \mathrm{Signal}_{t} \\[4pt]
    \text{Sell}, & \text{if } \mathrm{MACD}_{t-1} \geq \mathrm{Signal}_{t-1}
                   \text{ and } \mathrm{MACD}_{t} < \mathrm{Signal}_{t}
  \end{cases}
  \label{eq:rule_a}
\end{equation}

\noindent\textbf{MACD zero-line crossover strategy}


Another approach considers the position of the MACD relative to the
zero line. A buy signal occurs when the MACD crosses above zero,
indicating that short-term momentum exceeds long-term momentum. A sell
signal occurs when the MACD crosses below zero. Trading signals are
defined as:
\begin{equation}
  \text{Signal} =
  \begin{cases}
    \text{Buy},  & \text{if } \mathrm{MACD}_{t-1} \leq 0
                   \text{ and } \mathrm{MACD}_{t} > 0 \\[4pt]
    \text{Sell}, & \text{if } \mathrm{MACD}_{t-1} \geq 0
                   \text{ and } \mathrm{MACD}_{t} < 0
  \end{cases}
  \label{eq:rule_b}
\end{equation}

\noindent\textbf{MACD divergence}

MACD divergence refers to a mismatch between the direction of price movement and the direction of the MACD line, as defined in Eq.~\eqref{eq:macd}. This mismatch is commonly interpreted as a warning that the current price trend may be losing momentum and that a potential reversal may occur. A bullish divergence occurs when prices form lower lows while the MACD line forms higher lows, suggesting that downward momentum is weakening and that a potential buy signal may emerge. Conversely, a bearish divergence occurs when prices form higher highs while the MACD line forms lower highs, indicating that upward momentum may be weakening and that a potential sell signal may emerge.

In this study, we focus on the signal-line crossover strategy as the foundational trading mechanism. Given its widespread adoption in technical analysis, this approach provides a standardized benchmark for evaluating the empirical performance of the proposed VP-MACD model. To maintain the internal validity of the comparative analysis, other trading rules, including zero-line crossovers and MACD divergence, are excluded from the initial scope. By isolating this specific rule, we can more effectively identify the behavioral characteristics of the indicators before addressing their inherent structural limitations in subsequent sections.



\subsection{Limitations of the Traditional MACD}
\label{sec:limitations}

As discussed in Section~\ref{sec:lit_macd_lim}, prior studies have identified several limitations of the traditional MACD framework, including signal lag, sensitivity to noisy price movements, and the exclusion of supplementary market information such as volume and intraday price structure. From a methodological perspective, these limitations directly motivate the design of the proposed VP-MACD model. Because both the MACD line and the signal line are derived from smoothed historical price series, the indicator may react to price movements with delay, causing trading signals to appear only after a trend has already begun and thereby reducing entry-timing accuracy. In addition, the traditional MACD is constructed solely from price-based moving averages and does not explicitly incorporate trading volume, even though volume provides useful information about the strength of market participation and can help distinguish stronger price movements from weaker or noisier fluctuations. Furthermore, the traditional MACD primarily relies on closing prices and does not capture intraday price structure, such as the relationship between open, high, low, and close prices, which may contain information about intraday momentum, price range, and market pressure that is not reflected in closing prices alone. 

These limitations motivate the development of the proposed VP-MACD model, which embeds volume, volatility, and intraday price structure into the MACD price input. In addition, the \(\lambda\)-adjusted entry rule is introduced to mitigate signal lag and improve entry timing.

\subsection{Volume-price-adjusted MACD}
\label{sec:vpmacd}

Building on these limitations, this study proposes a
volume-price-adjusted MACD (VP-MACD), which incorporates volume, price
volatility, and structural price information into the conventional MACD
framework. Specifically, we define an adjusted price $P_{t}^{*}$ as a
volume-weighted adjusted price that integrates both volatility and
intraday price movement: 

\begin{equation}
P_t^{*} =
\frac{
\sum_{i=t-N}^{t-1} P_i \cdot \mathrm{Volume}_i \cdot \sigma_i \cdot r_i
}{
\sum_{i=t-N}^{t-1} \mathrm{Volume}_i
}
\label{eq:adj_price}
\end{equation}
where $t$ denotes the current trading day, $i$ indexes historical observations within the lookback window, and $N$ represents the number of past trading days included in the rolling calculation. In addition, $P_i$ denotes the closing price on day $i$, $\mathrm{Volume}_i$ denotes the trading volume on day $i$, $\sigma_i$ captures the relative volatility component, and $r_i$ measures intraday price structure.

The volume term serves as the primary weighting mechanism, while the volatility component $\sigma_i$ captures the magnitude of price variation within each period. The relative price measure $r_i$ further provides information on the directional strength of intraday price movement. Together, these components allow the adjusted price $P_t^*$ to incorporate trading intensity, price variability, and directional price structure into the conventional MACD framework.

The volatility term is defined as:
\begin{equation}
  \sigma_{i} = \frac{\mathrm{STD}\!\left(P_{i}^{\mathrm{High}} - P_{i}^{\mathrm{Low}}\right)}{P_i^{\mathrm{Close}}}
  \label{eq:sigma}
\end{equation}


\noindent
where $P_i^{\mathrm{High}}$, $P_i^{\mathrm{Low}}$, and $P_i^{\mathrm{Close}}$ denote the high, low, and closing prices on day $i$, respectively. The numerator measures the variation in the daily high--low price range, while dividing by $P_i^{\mathrm{Close}}$ normalizes the volatility measure and makes it scale-adjusted across different price levels.

and the relative price measure is defined as:
\begin{equation}
  r_{i} = \frac{\left|P_{i}^{\mathrm{Close}} - P_{i}^{\mathrm{Open}}\right|}
               {P_{i}^{\mathrm{High}} - P_{i}^{\mathrm{Low}}}
  \label{eq:ri}
\end{equation}
\noindent
where \(P_i^{\mathrm{Open}}\) denotes the opening price on day \(i\).

The relative price measure $r_{i}$ is designed to capture the
directional strength of intraday price movement by normalizing the
difference between closing and opening prices relative to the total
price range. While volatility $\sigma_{i}$ reflects the magnitude of
price fluctuations, it does not distinguish between different
candlestick structures that may carry distinct market implications.


\begin{figure}[!htbp]
    \centering
    \includegraphics[width=0.5\linewidth]{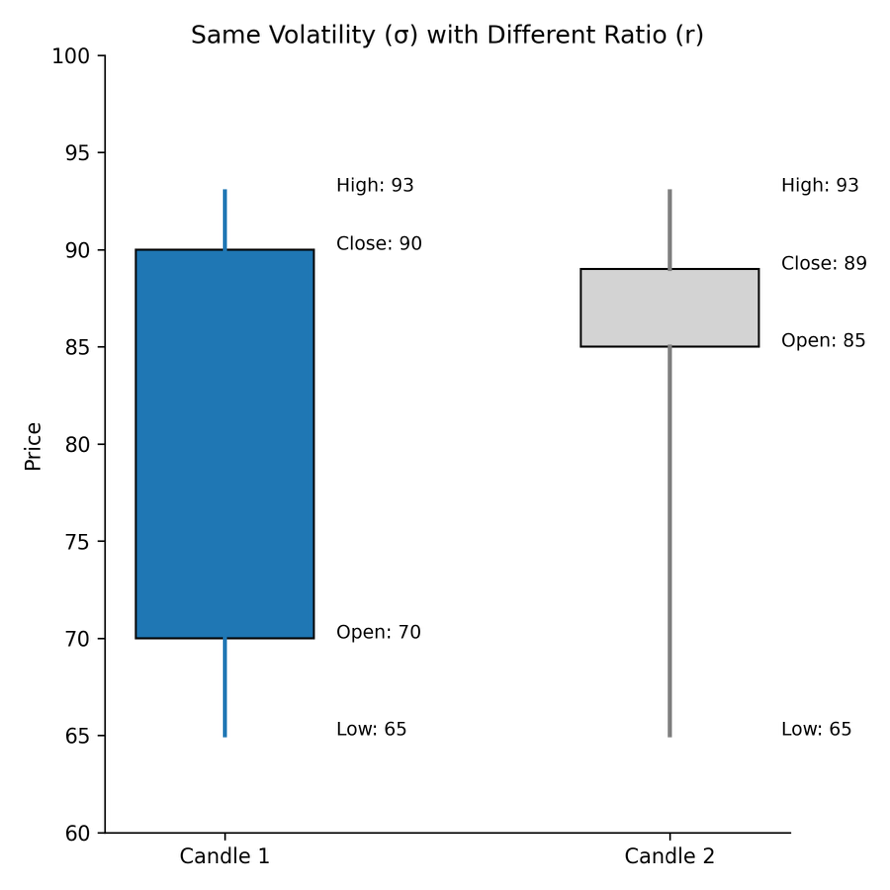}
    \caption{Illustration of two candlesticks with the same volatility measure ($\sigma_i$) but different directional strength ($r_i$).}
    \label{fig:candlestick}
\end{figure}

As illustrated in Fig.~\ref{fig:candlestick}, two price movements may
exhibit identical volatility but represent fundamentally different
market conditions. In the left candlestick, the large body indicates a
strong directional move from open to close, suggesting sustained buying
pressure. In contrast, the right candlestick, despite having the same
high-low range, exhibits a smaller body with longer wicks, indicating
greater intraday uncertainty and weaker directional conviction.

By incorporating $r_{i}$ into the adjusted price, the proposed method
differentiates between strong trends and noisy fluctuations, leading to
more informative signal generation. This allows the model to jointly
account for both the magnitude of price variation (captured by
$\sigma_{i}$) and the directional intensity of price movement (captured
by $r_{i}$).

Using the adjusted price $P_{t}^{*}$, the VP-MACD indicator is
constructed in a manner analogous to the traditional MACD:
\begin{equation}
  \mathrm{VP}\text{-}\mathrm{MACD}_{t}
    = \mathrm{EMA}_{12}(P_{t}^{*}) - \mathrm{EMA}_{26}(P_{t}^{*})
  \label{eq:vpmacd}
\end{equation}

and the signal line is defined as:
\begin{equation}
  \mathrm{Signal}_{t} = \mathrm{EMA}_{9}(\mathrm{VP}\text{-}\mathrm{MACD}_{t})
  \label{eq:vpsignal}
\end{equation}

To further address the lagging issue in signal generation, we introduce a sensitivity parameter \(\lambda\) to modify the entry rule. Instead of strictly relying on the standard crossover condition, the buy signal is adjusted to allow earlier market participation. In particular, the buy signal is triggered when the VP-MACD line exceeds the signal line scaled by a factor \(\lambda\), where \(\lambda \in (0.8, 1)\).

The interval \((0.8, 1)\) is predefined as a practical calibration range rather than derived from a theoretical restriction. Since \(\lambda < 1\) lowers the effective signal-line threshold, it allows the strategy to enter the market before a conventional crossover occurs. At the same time, restricting \(\lambda\) to be greater than 0.8 prevents the entry condition from being excessively relaxed, thereby reducing the risk of premature signals caused by market noise. The resulting \(\lambda\)-adjusted trading rule is defined as:
\begin{equation}
\mathrm{Signal}_{t}^{\lambda} =
\begin{cases}
\mathrm{Buy}, & \text{if } \mathrm{VP\text{-}MACD}_{t-1} \leq \lambda \mathrm{Signal}_{t-1}
\text{ and } \mathrm{VP\text{-}MACD}_{t} > \lambda \mathrm{Signal}_{t}, \\[4pt]
\mathrm{Sell}, & \text{if } \mathrm{VP\text{-}MACD}_{t-1} \geq \mathrm{Signal}_{t-1}
\text{ and } \mathrm{VP\text{-}MACD}_{t} < \mathrm{Signal}_{t}.
\end{cases}
\label{eq:lambda_rule}
\end{equation}
%

\subsection{Backtesting Framework and Evaluation Metrics}
\label{sec:backtest}

This section outlines the backtesting framework and evaluation methodology used to assess the performance of the proposed VP-MACD strategy in comparison with benchmark strategies. Specifically, we describe the data sources, trading rules, portfolio simulation setup, and evaluation metrics. For consistency throughout the empirical analysis, this study evaluates three strategies: Strategy A, the traditional MACD crossover strategy; Strategy B, the MACD+$\lambda$ strategy, which applies a $\lambda$-adjusted entry rule to the traditional MACD signal; and Strategy C, the proposed VP-MACD strategy, which applies its own separately optimized $\lambda$-adjusted entry rule. Multiple performance metrics are then employed to compare these strategies across different market conditions.

\subsubsection{Data Description}
\label{sec:data}

This study uses daily market data from three major U.S.\ equity
indices: the S\&P~500 (SPY), the NASDAQ~100 (QQQ), and the Dow Jones
Industrial Average (DIA). These indices represent different segments of
the U.S.\ equity market, ranging from diversified large-cap stocks,
growth-oriented technology firms, and blue-chip companies.

Structurally, the S\&P~500 and NASDAQ~100 employ a
market-capitalization-weighted methodology, whereas the Dow Jones
Industrial Average is price-weighted, resulting in differences in index
construction and price dynamics. While the NASDAQ~100 is characterized
by its high concentration in the technology sector, the S\&P~500
provides broader diversification across its 11 sectors. All three
indices exhibit high liquidity with substantial trading volume, which
ensures reliable pricing and reduces market noise, leading to more
stable and realistic backtesting results. Despite some overlap in
constituents, the divergent sectoral compositions and volatility
characteristics of these indices facilitate a comprehensive evaluation
of the proposed strategy under varying market conditions.

The data are obtained from Yahoo Finance and include five raw market features: Open, High, Low, Close, and trading volume. The sample period spans from 2018 to February 2026. Because the data are reported at the daily trading frequency, weekends and non-trading days are excluded from the sample. To mitigate overfitting, the dataset is divided into a training period from 2018 to 2022 and a testing period from 2023 to February 2026. The training sample contains 1,259 trading-day observations, while the testing sample contains 791 trading-day observations for each instrument. Model parameters are calibrated on the training set and evaluated on the testing set. This split ensures that the evaluation reflects out-of-sample performance.

\subsubsection{Backtesting Setup}
\label{sec:setup}

The backtesting simulation is initialized with a starting capital of \$100,000. To maintain simplicity in the baseline evaluation, no leverage is applied, and positions are fully invested upon signal execution with a minimum trading unit of one share. We adopt a long-only trading strategy: a position is opened when a buy signal is generated and closed when a sell signal occurs. After exiting a position, the strategy waits for the next valid buy signal before re-entering the market. Since the MACD indicator is computed based on daily closing prices, trades cannot be executed at the signal generation time. To avoid look-ahead bias, all trades are executed at the opening price of the next trading day following a signal. This ensures that the simulation reflects a realistic execution environment where signals are processed after market close. Transaction costs are incorporated using a one-way cost assumption of 4 bps per trade, consisting of 3 bps in trading cost and 1 bp in slippage. The cost is applied at both entry and exit, resulting in a total round-trip cost of 8 bps per completed trade.

To evaluate the impact of the parameter $\lambda$, we conduct a grid
search over the range $\lambda \in [0.8, 1.0]$ at an increment of 0.02.
The optimal $\lambda$ is determined separately for each strategy (MACD+$\lambda$ and VP-MACD) and for each index using the training dataset. The selected strategy-specific and index-specific $\lambda$  values are then fixed and applied to the testing dataset, allowing for different optimal parameter values across markets and different strategies.

\subsubsection{Evaluation Metrics}
\label{sec:metrics}

To comprehensively evaluate the performance of the proposed VP-MACD
strategy relative to the benchmark models, a set of performance metrics
is employed to assess profitability, risk exposure, trading efficiency,
and statistical significance from multiple perspectives. These metrics
are selected to provide both economic interpretation and practical
trading insights.

First, trading activity is measured by the total number of trades,
defined as:
\begin{equation}
  \text{Total Trades} = N
  \label{eq:trades}
\end{equation}
where each completed buy--sell pair is counted as one trade. This
metric reflects the frequency of trading signals and provides insight
into the strategy's level of market participation. A higher number of
trades may indicate more opportunities but can also imply increased
exposure to noise and transaction costs (if considered in practice).

Second, the win ratio evaluates the consistency of the strategy in
generating profitable trades:
\begin{equation}
  \text{Win Ratio} = \frac{N_{\mathrm{win}}}{N}
  \label{eq:winratio}
\end{equation}
where $N_{\mathrm{win}}$ denotes the number of profitable trades and
$N$ is the total number of trades. A higher win ratio indicates that
the strategy produces more consistent trading outcomes, although it
does not account for the magnitude of gains and losses.

Third, the profit and loss (PnL) ratio captures the relationship
between average gains and average losses:
\begin{equation}
  \text{PnL Ratio} = \frac{\mathrm{Gains}/N_{\mathrm{win}}}{\mathrm{Losses}/N_{\mathrm{Loss}}}
  \label{eq:pnlratio}
\end{equation}
where $N_{\mathrm{Loss}}$ represents the number of losing trades. This
metric evaluates the risk-reward profile of the strategy, indicating
whether gains from profitable trades are sufficient to compensate for
losses. In particular, a PnL ratio greater than 1 implies that the
average gain exceeds the average loss, meaning that profits are
sufficient to cover losses on a per-trade basis. Even with a moderate
or low win ratio, a strategy can remain profitable if the PnL ratio is
sufficiently high. Importantly, a strategy with a win ratio below 50\%
can still achieve positive overall profitability, as long as the
magnitude of gains from winning trades significantly exceeds the losses
from losing trades.

Fourth, the expectancy metric is used to measure the expected profit
per trade by combining both the win ratio and the PnL ratio:
\begin{equation}
  \text{Expectancy} = \text{Win Ratio} \times \overline{G}
                    - (1 - \text{Win Ratio}) \times \overline{L}
  \label{eq:expectancy}
\end{equation}
where $\overline{G}$ and $\overline{L}$ denote average gain and average
loss per trade, respectively. This metric reflects the average return
per trade and provides a direct measure of the strategy's
profitability. A positive expectancy indicates that the strategy is
profitable on average, while a negative value suggests that losses
outweigh gains over time. Importantly, expectancy integrates the
frequency of winning trades and the magnitude of gains and losses,
offering a more comprehensive assessment than win ratio or PnL ratio
alone.

In addition, the total profit and loss (PnL) is calculated as:
\begin{equation}
  \text{Total PnL} = V_{T} - V_{0}
  \label{eq:totalpnl}
\end{equation}
where $V_{T}$ is the final portfolio value and $V_{0}$ is the initial
capital. This metric captures the absolute profitability of the
strategy over the backtesting period and reflects its overall economic
value.

To evaluate risk-adjusted performance, the Sharpe ratio is employed:
\begin{equation}
  \text{Sharpe Ratio} = \frac{\mathbb{E}[R_{t}]}{\sigma(R_{t})} \times \sqrt{252}
  \label{eq:sharpe}
\end{equation}
where $R_{t}$ denotes daily returns and $\sigma(R_{t})$ is their
standard deviation. The Sharpe ratio evaluates how efficiently the
strategy generates returns relative to the level of risk undertaken,
allowing for comparison across strategies with different volatility
profiles.

Furthermore, the maximum drawdown (MDD) is used to quantify downside
risk:
\begin{equation}
  \mathrm{MDD} = \max_{t}
    \left(\frac{\mathrm{Peak}_{t} - \mathrm{Value}_{t}}{\mathrm{Peak}_{t}}\right)
  \label{eq:mdd}
\end{equation}
where $\mathrm{Peak}_{t} = \max_{s \leq t} \mathrm{Value}_{s}$ denotes
the maximum portfolio value observed up to time $t$. This metric
measures the largest loss from peak to a subsequent trough in portfolio
value, reflecting the worst-case loss scenario faced by investors. It
is particularly important for assessing the robustness and risk
tolerance of the strategy in adverse market conditions.

Finally, to assess whether the performance improvement of the proposed
strategy is statistically significant, a one-sided $t$-test is
conducted on the mean returns across different strategy pairs, including
comparisons between the traditional MACD, the MACD+$\lambda$,
and the proposed VP-MACD model. This stepwise comparison enables us to
isolate the incremental contribution of the $\lambda$-adjustment and
the VP enhancement. The hypotheses are defined as:
\begin{equation}
  H_{0}: \mu_{\mathrm{new}} \leq \mu_{\mathrm{old}}
  \qquad
  H_{A}: \mu_{\mathrm{new}} > \mu_{\mathrm{old}}
  \label{eq:hyp}
\end{equation}

Specifically, three pairwise comparisons are performed: (i) traditional
MACD vs.\ MACD+$\lambda$, (ii) MACD+$\lambda$
vs.\ VP-MACD, and (iii) traditional MACD vs.\ VP-MACD. A one-sided
test is adopted since the objective is to verify performance
enhancement rather than merely detect differences. Rejection of the
null hypothesis indicates that the proposed method delivers
statistically superior mean return performance.

Given that financial return differences may exhibit heteroskedasticity and serial dependence, HAC t-test with Newey-West adjusted standard errors are further employed. In addition, a blocked circular bootstrap is conducted as a non-parametric robustness check.

Collectively, these metrics jointly provide a comprehensive evaluation
framework that captures trading effectiveness, profitability, risk
exposure, and statistical robustness, enabling a thorough comparison
between the proposed model and benchmark strategies.

\section{Results}
\label{sec:results}

\subsection{Trading Characteristics}
\label{sec:training}
To compare the three strategies, the sensitivity parameter \(\lambda\) is optimized separately for the MACD+$\lambda$ and VP-MACD strategies for each index using the training sample. A grid search is conducted over the predefined range \(\lambda \in [0.8, 1.0]\) with an increment of 0.02, yielding eleven candidate values for each strategy-index combination. The optimal \(\lambda\) is selected through a multi-criteria evaluation of six trading performance metrics: win ratio, total profit and loss (PnL), PnL ratio, Sharpe ratio, maximum drawdown, and expected value. Rather than maximizing a single metric, the selection balances return generation, risk-adjusted performance, downside risk, and per-trade profitability. Specifically, preference is given to \(\lambda\) values that improve Sharpe ratio, PnL ratio, and expected value without materially increasing maximum drawdown or relying on unstable changes in total PnL. The full metric profiles across  $\lambda$ values for the MACD+$\lambda$ and VP-MACD strategies are presented in Figures ~\ref{fig:spy}--\ref{fig:dia} and Figures.~\ref{fig:spy_vp_macd}--\ref{fig:dia_vp_macd}, respectively.

For the  MACD+$\lambda$ strategy, the performance metrics exhibit notably different patterns as  $\lambda$
varies, ultimately leading to distinct optimal $\lambda$ selections that reflect each index's underlying market characteristics. 
For SPY, while a higher $\lambda$ generally yields greater absolute returns along with higher PnL ratio and expected value, these gains are accompanied by increased volatility and drawdown. Given that the Sharpe ratio reaches a favorable balance around  $\lambda$= 0.90 without materially worsening downside risk, this value is selected as optimal. 
For QQQ, performance metrics exhibit greater instability across the candidate  $\lambda$ values, reflecting the index's comparatively higher volatility and sensitivity to parameter choice. Although the Sharpe ratio peaks near  $\lambda=0.84$, metrics such as expected value and PnL ratio do not improve consistently beyond this point, suggesting that a lower $\lambda$ enhances responsiveness to short-term price movements without overfitting to noise. Accordingly, $\lambda=0.84$ is selected.
For DIA, the metric profiles display a clearer and more monotonic improvement as $\lambda$ increases. The PnL ratio and expected value rise steadily, the Sharpe ratio improves markedly near $\lambda=0.92$, and maximum drawdown remains relatively contained throughout. This pattern indicates that a higher $\lambda$ strengthens signal quality without disproportionately elevating risk, leading to the selection of $\lambda=0.92$

For the VP-MACD strategy, a different pattern emerges across the three indices compared to the MACD+$\lambda$ MACD. The incorporation of volume, volatility, and price structure information shifts the optimal $\lambda$ values, reflecting the changed informational basis of the indicator. For SPY and DIA, the optimal $\lambda$ values are slightly lower at 0.88 and 0.86, respectively, suggesting that the additional market information allows the indicator to accommodate stronger sensitivity adjustment without introducing excessive noise, thereby generating more responsive signals. In contrast, for QQQ, the optimal $\lambda$ increases substantially to 0.98, indicating that in a highly volatile market environment, a lower $\lambda$ risks amplifying noise and destabilizing signals. A higher $\lambda$ corresponding to a more conservative crossover condition and smoother signal generation, is therefore preferred to prevent the strategy from overreacting to short-term price fluctuations.

Overall, these results suggest that the optimal $\lambda$ is jointly determined by market volatility and the informational structure of the indicator. In relatively stable markets, the incorporation of additional information such as volume, volatility, and price structure allows for stronger sensitivity adjustment, whereas more volatile markets require a more conservative, higher $\lambda$ setting to maintain signal robustness. Table \ref{tab:lambda} summarizes the optimal $\lambda$ selections for both the $\lambda-$adjusted MACD and VP-MACD strategies across all three indices.

\begin{table}[H]
\centering
\small
\caption{Optimal $\lambda$ selections for the MACD+$\lambda$ and VP-MACD strategies across indices.}
\label{tab:lambda}
\begin{tabular}{lccc}
\hline
\textbf{Index} & \textbf{MACD+$\lambda$} & \textbf{VP-MACD} & \textbf{Primary Selection Criterion} \\
\hline
SPY & 0.90 & 0.88 & Sharpe ratio \\
QQQ & 0.84 & 0.98 & Sharpe ratio \\
DIA & 0.92 & 0.86 & Sharpe ratio + Expected value \\
\hline
\end{tabular}
\end{table}

\subsection{Performance Evaluation}
\label{sec:eval}
This section evaluates the performance of the three trading strategies using the testing data from January 1, 2023 to February 28, 2026. The training data are used only for parameter calibration, including the selection of the sensitivity parameter $\lambda$, whereas the testing data are used for performance evaluation.

As defined in the backtesting framework in Section~\ref{sec:backtest}, the three strategies are Strategy A, the baseline MACD crossover strategy; Strategy B, the MACD+$\lambda$ strategy, which applies an optimized $\lambda$-adjusted entry rule to the traditional MACD signal; and Strategy C, the proposed VP-MACD strategy, which applies its own separately optimized $\lambda$-adjusted entry rule. For both Strategy B and Strategy C, the optimal $\lambda$ values are calibrated separately on the training data for each index before being applied to the testing period.

\subsubsection{Effect of $\lambda$: Strategy B vs.\ Strategy A}

Across all three indices, Strategy B improves the win ratio relative to
Strategy A, indicating that the introduction of $\lambda$ enhances
signal quality by filtering out weaker and noisier trading signals.
This suggests that $\lambda$ acts as a smoothing or filtering mechanism
that prioritizes stronger trends.

For SPY and QQQ, Strategy B also improves risk-adjusted performance, as
reflected in higher Sharpe ratios and PnL ratios. This indicates that
removing low-quality signals leads to more consistent returns. However, this improvement is not uniform across all markets. For DIA,
although the win ratio increases, the total PnL decreases and becomes
negative. This implies that some profitable trades are filtered out
along with noisy signals. In relatively less trending or
slower-moving markets like DIA, aggressive filtering may eliminate not
only false signals but also smaller yet meaningful opportunities.

Taken together, $\lambda$ improves trade quality but introduces a trade-off
by reducing trading frequency and potentially missing profitable
signals. 

\subsubsection{Impact of Additional Information: Strategy C vs.\ Strategy B}

Across all three indices, Strategy C improves key risk-control and trade-quality metrics relative to Strategy B. In particular, the expected value increases for all three indices, while the maximum drawdown is reduced in each case. The improvement in expected value is especially notable: it increases by more than 200\% for SPY, by approximately 57\% for QQQ, and turns from negative to positive for DIA.

For QQQ, Strategy C further improves both profitability and risk-adjusted performance. The higher Sharpe ratio, improved expected value, and reduced maximum drawdown suggest that incorporating volume, volatility, and price-structure information helps confirm trend strength and filter out false breakouts. In a highly volatile market such as QQQ, this additional information appears to enhance signal reliability and improve overall performance.

For SPY, Strategy C improves some trade-quality metrics, including the PnL ratio and expected value, but its total PnL does not exceed that of Strategy B. The number of trades decreases sharply from 36 to 6. Although the PnL ratio is substantially higher, the reduced trading frequency limits total profitability. This may be explained by the highly liquid and relatively stable nature of SPY, where price information already captures much of the market trend. In such an environment, adding additional volume, volatility, and price-structure information may remove valid trading opportunities, so the gains in trade quality do not fully translate into higher total profitability.

For DIA, Strategy C substantially outperforms Strategy B. It not only reverses the negative total PnL observed under Strategy B, but also improves most other performance metrics. This suggests that incorporating volume, volatility, and price-structure information helps recover useful signals that are not captured by the $\lambda$ adjustment alone. In markets with weaker or less persistent trends, additional contextual information becomes important for distinguishing between noise and genuine trading opportunities.

In this comparison, the benefit of Strategy C is strongest for QQQ and DIA, while the SPY result is more mixed because the number of VP-MACD trades is very small. These results suggest that VP-MACD improves signal quality and risk control, but its effect on total profitability depends on market conditions, particularly volatility, trend strength, and trading frequency.

\begin{table}[!htbp]
\centering
\caption{Comparison of performance metrics across trading strategies for SPY.}
\label{tbl:spy}
\small
\resizebox{\textwidth}{!}{
\begin{tabular}{lccccccc}
\toprule
\textbf{Strategy} & \textbf{Trades} & \textbf{Win Rate} & \textbf{Total PnL} & \textbf{PnL Ratio} & \textbf{Sharpe} & \textbf{Max DD} & \textbf{Expectancy} \\
\midrule
Baseline MACD & 40 & 50.00\% & \$13,612.34 & 1.39 & 0.47 & $-13.78\%$ & \$19.39 \\
MACD ($\lambda = 0.90$) & 36 & 55.56\% & \$52,688.94 & 2.98 & 1.35 & $-8.17\%$ & \$121.04 \\
\textbf{VP-MACD ($\lambda = 0.88$)} & \textbf{6} & \textbf{50.00\%} & \textbf{\$27,543.78} & \textbf{8.61} & \textbf{0.96} & \textbf{$-6.33\%$} & \textbf{\$365.72} \\
\bottomrule
\end{tabular}
}
\end{table}

\begin{table}[!htbp]
\centering
\caption{Comparison of performance metrics across trading strategies for QQQ.}
\label{tbl:qqq}
\small
\resizebox{\textwidth}{!}{
\begin{tabular}{lccccccc}
\toprule
\textbf{Strategy} & \textbf{Trades} & \textbf{Win Rate} & \textbf{Total PnL} & \textbf{PnL Ratio} & \textbf{Sharpe} & \textbf{Max DD} & \textbf{Expectancy} \\
\midrule
Baseline MACD & 35 & 45.71\% & \$16,823.12 & 1.56 & 0.45 & $-22.84\%$ & \$17.14 \\
MACD ($\lambda = 0.84$) & 31 & 61.29\% & \$63,321.06 & 1.89 & 1.13 & $-11.94\%$ & \$76.83 \\
\textbf{VP-MACD ($\lambda = 0.98$)} & \textbf{21} & \textbf{80.95\%} & \textbf{\$88,324.48} & \textbf{1.72} & \textbf{1.51} & \textbf{$-7.41\%$} & \textbf{\$120.26} \\
\bottomrule
\end{tabular}
}
\end{table}

\begin{table}[!htbp]
\centering
\caption{Comparison of performance metrics across trading strategies for DIA.}
\label{tbl:dia}
\small
\resizebox{\textwidth}{!}{
\begin{tabular}{lccccccc}
\toprule
\textbf{Strategy} & \textbf{Trades} & \textbf{Win Rate} & \textbf{Total PnL} & \textbf{PnL Ratio} & \textbf{Sharpe} & \textbf{Max DD} & \textbf{Expectancy} \\
\midrule
Baseline MACD & 40 & 40.00\% & \$1,236.96 & 1.55 & 0.09 & $-13.95\%$ & \$2.09 \\
MACD ($\lambda = 0.92$) & 34 & 50.00\% & $-\$6,862.00$ & 0.75 & $-0.20$ & $-14.19\%$ & $-\$12.55$ \\
\textbf{VP-MACD ($\lambda = 0.86$)} & \textbf{13} & \textbf{69.23\%} & \textbf{\$29,592.17} & \textbf{3.72} & \textbf{1.08} & \textbf{$-6.20\%$} & \textbf{\$226.87} \\
\bottomrule
\end{tabular}
}
\end{table}

\subsubsection{Overall Comparison and Trade-off}

Compared to the baseline Strategy A, Strategy C improves all key performance metrics, including win ratio, total PnL, PnL ratio, Sharpe ratio, maximum drawdown, and expected value. From an economic perspective, these improvements are meaningful because they reflect stronger profitability, better risk control, and improved trade-level performance. This demonstrates that integrating the $\lambda$ adjustment with volume, volatility, and price-related information enhances both profitability and risk-adjusted performance relative to the traditional MACD strategy.

Both Strategy B and Strategy C result in fewer trades compared to Strategy A, reflecting the filtering effect introduced by the $\lambda$ adjustment and the additional information incorporated in VP-MACD. While this improves signal quality and reduces false positives, it also introduces the risk of missing profitable opportunities. From a market perspective, this highlights a fundamental trade-off between signal precision and market responsiveness. In more volatile markets, stronger filtering appears more effective because additional information helps distinguish meaningful price movements from short-term noise. In relatively stable markets, excessive filtering may remove valid trading opportunities and reduce trading frequency, thereby limiting the overall benefit of the enhanced strategy.

Collectively, Strategy B improves signal quality by adjusting the responsiveness of MACD through $\lambda$, while Strategy C further extends the framework by incorporating additional market information. These findings suggest that the proposed VP-MACD model is most effective when the added information improves signal selection without excessively reducing the strategy's ability to respond to valid trading opportunities.

\subsection{Statistical Significance of Performance Improvements} 
\label{sec:hyptest}

The performance results in Section~4.2 suggest that the proposed VP-MACD strategy provides economically meaningful improvements in profitability, risk control, and trade-level performance. To further examine whether these improvements are statistically significant, we conduct three complementary tests on the differences in daily returns between strategies over the testing period from January 1, 2023 to February 28, 2026. Consistent with the backtesting design, Strategy B and Strategy C each use their own optimized $\lambda$ values, calibrated separately on the training data before being applied to the testing period.

First, we use one-sided $t$-tests to examine whether the mean return difference between two strategies is significantly greater than zero. Second, because financial return series may exhibit heteroskedasticity and autocorrelation, we apply Newey--West adjusted $t$-tests as a robustness check. Third, we conduct circular block bootstrap tests to provide a non-parametric validation of the results without relying heavily on normality assumptions. Specifically, the bootstrap procedure uses 1,000 resamples with a block length of five consecutive daily returns.

Following the pairwise comparison framework defined in Section~3.5.3, this section evaluates the same three pairwise comparisons: Pair 1 compares Strategy B with Strategy A; Pair 2 compares Strategy C with Strategy A; and Pair 3 compares Strategy C with Strategy B. The hypotheses are defined as in Eq.~(\ref{eq:hyp}), where $\mu$ represents the mean daily return of each strategy. A positive $t$-statistic value indicates that the first strategy in each pair outperforms the corresponding benchmark strategy.

\begin{table}[!htbp]
\centering
\caption{One-sided $t$-test results for pairwise strategy comparisons.}
\label{tbl:ttest}
\small
\resizebox{\textwidth}{!}{
\begin{tabular}{lcccccc}
\toprule
\textbf{Comparison} & \textbf{SPY ($t$)} & \textbf{SPY ($p$)} & \textbf{QQQ ($t$)} & \textbf{QQQ ($p$)} & \textbf{DIA ($t$)} & \textbf{DIA ($p$)} \\
\midrule
Pair 1: MACD+$\lambda$ vs. MACD & 1.7171 & 0.0432$^{*}$ & 1.6653 & 0.0481$^{*}$ & $-0.6313$ & 0.0736 \\
Pair 2: VP-MACD vs. MACD & 0.4971 & 0.3096 & 1.6586 & 0.0488$^{*}$ & 1.2804 & 0.1004 \\
Pair 3: VP-MACD vs. MACD+$\lambda$ & $-0.8857$ & 0.8120 & 0.3107 & 0.3781 & 1.7065 & 0.0442$^{*}$ \\
\bottomrule
\end{tabular}
}
\vspace{0.3em}
\begin{minipage}{0.95\textwidth}
\footnotesize
\textit{Note.} $^{*}$ denotes statistical significance at the 5\% level ($p < 0.05$).
\end{minipage}
\end{table}

\begin{table}[!htbp]
\centering
\caption{Newey--West adjusted $t$-test results for pairwise strategy comparisons.}
\label{tbl:nwttest}
\small
\resizebox{\textwidth}{!}{
\begin{tabular}{lcccccc}
\toprule
\textbf{Comparison} & \textbf{SPY ($t$)} & \textbf{SPY ($p$)} & \textbf{QQQ ($t$)} & \textbf{QQQ ($p$)} & \textbf{DIA ($t$)} & \textbf{DIA ($p$)} \\
\midrule
Pair 1: MACD+$\lambda$ vs. MACD & 1.8983 & 0.0288$^{*}$ & 1.6931 & 0.0452$^{*}$ & $-0.6593$ & 0.7452 \\
Pair 2: VP-MACD vs. MACD & 0.5307 & 0.2978 & 1.7229 & 0.0425$^{*}$ & 1.2964 & 0.0974 \\
Pair 3: VP-MACD vs. MACD+$\lambda$ & $-0.9922$ & 0.8394 & 0.3487 & 0.3637 & 1.7894 & 0.0368$^{*}$ \\
\bottomrule
\end{tabular}
}
\vspace{0.3em}
\begin{minipage}{0.95\textwidth}
\footnotesize
\textit{Note.} $^{*}$ denotes statistical significance at the 5\% level ($p < 0.05$).
\end{minipage}
\end{table}

The standard one-sided $t$-test results are reported in Table \ref{tbl:ttest}. For SPY, Pair 1 is statistically significant ($p = 0.0432$), indicating that Strategy B, the MACD+$\lambda$ strategy, improves performance relative to Strategy A, the baseline MACD strategy. For QQQ, both Pair 1 ($p = 0.0481$) and Pair 2 ($p = 0.0488$) are significant, suggesting that both the $\lambda$ adjustment and the proposed VP-MACD framework improve performance relative to the baseline MACD strategy. For DIA, Pair 3 is significant ($p = 0.0442$), indicating that Strategy C, the VP-MACD strategy, significantly outperforms Strategy B, the MACD+$\lambda$ strategy.

Table~\ref{tbl:nwttest} reports the Newey--West adjusted $t$-test results. The significance pattern remains largely consistent with the standard $t$-tests. Pair 1 remains significant for SPY ($p = 0.0288$) and QQQ ($p = 0.0452$), Pair 2 remains significant for QQQ ($p = 0.0425$), and Pair 3 remains significant for DIA ($p = 0.0368$). This consistency suggests that the main findings are robust after adjusting for potential heteroskedasticity and autocorrelation in the return series.

To further validate the results, we also conduct circular block bootstrap tests. Figures~8--10 confirm the key significant findings: Pair 1 is significant for SPY ($p = 0.024$), Pair 2 is significant for QQQ ($p = 0.039$), and Pair 3 is significant for DIA ($p = 0.028$). The bootstrap distributions are shifted to the right of zero in these cases, indicating that the mean return differences are positive.

The hypothesis testing results therefore support a two-step interpretation of the strategy improvements. First, the $\lambda$ adjustment improves the traditional MACD strategy in selected markets, as shown by the significance of Pair 1 for SPY and QQQ. Second, the additional volume, volatility, and price-structure information in VP-MACD provides incremental value beyond the baseline MACD for QQQ and beyond MACD+$\lambda$ for DIA, as shown by the significance of Pair 2 for QQQ and Pair 3 for DIA. These results suggest that the proposed enhancements improve performance across the three indices in different ways, although their effectiveness remains market-dependent.

\section{Conclusions}
\label{sec:conclusions}

This study proposes a volume-price-adjusted MACD (VP-MACD) framework, together with a sensitivity parameter, $\lambda$, to address the lagging behavior and false signals of the traditional MACD model.

A key component of the proposed framework is the determination of optimal $\lambda$ values for both the MACD+$\lambda$ strategy and the VP-MACD strategy. Using the training period from 2018 to 2022, the optimal $\lambda$ values are selected separately for each strategy and each index, and are then applied to the testing period from 2023 to February 2026. This strategy and index specific design is motivated by differences in market structure. SPY is characterized by the highest liquidity and moderate volatility, QQQ by high liquidity and the highest volatility, and DIA by relatively lower liquidity and lower volatility. The selected $\lambda$ values suggest that markets with different volatility and trend characteristics may require different levels of signal responsiveness.

The multi-metric evaluation framework shows that the VP-MACD model generally outperforms the baseline MACD model in terms of profitability, risk-adjusted return, and downside-risk control across the three indices. In particular, expectancy, which measures the average profit or loss per trade and captures the combined effect of win ratio and PnL ratio, emerges as an especially important indicator of long-run trading viability. The results show that expectancy improves substantially under VP-MACD relative to the baseline MACD, increasing by approximately 7 to 109 times across the three indices. These improvements indicate that the proposed framework delivers economically meaningful gains in trade-level performance over the traditional MACD approach.
In addition, the comparison between VP-MACD and MACD+$\lambda$ shows that the added volume, volatility, and price-structure information can provide value beyond the $\lambda$ adjustment alone, especially for DIA, where VP-MACD reverses the negative performance of MACD+$\lambda$ and produces stronger overall results.

The statistical tests further support the economic findings. The one-sided $t$-tests, Newey–West adjusted $t$-tests, and bootstrap results provide consistent evidence that the proposed enhancements generate statistically significant improvements in selected markets. Specifically, MACD+$\lambda$ shows significant improvement for SPY, VP-MACD significantly improves performance for QQQ relative to the baseline MACD, and VP-MACD significantly outperforms MACD+$\lambda$ for DIA. These findings should therefore be interpreted as evidence that VP-MACD improves signal quality most effectively when the additional information helps identify meaningful opportunities without excessively reducing valid trades. More broadly, this market-dependent result is consistent with recent evidence that volatility dynamics and hedging effectiveness vary across market regimes, suggesting that trading indicators should be calibrated with attention to changing market conditions \citep{cheng2026regime}.

An important implication of the analysis is the trade-off between signal selectivity and trading frequency. The VP-MACD framework generally produces fewer trades than the baseline MACD, suggesting that the added volume, volatility, and price-structure information helps filter weaker or noisier signals. This selectivity can improve trade quality, risk control, and expected value, but it may also reduce the number of realized trades and limit statistical power when the sample of executed trades is small. These findings should therefore be interpreted as evidence that VP-MACD improves signal quality most effectively when the additional information helps identify meaningful opportunities without excessively reducing valid trades. Although the testing period contains approximately 800 daily observations, the number of executed trades can be limited under stricter signal-filtering rules, especially for VP-MACD. Future research may extend the testing horizon, use higher-frequency data, or evaluate the VP-MACD framework across a broader set of markets to obtain more trading observations and further assess the statistical significance and robustness of the proposed strategy. In addition, although this study focuses on the U.S. equity market, the VP-MACD framework could be extended to other equity markets, such as China and Japan, as well as to other asset classes, including foreign currencies, commodities, and cryptocurrencies.

\section*{Statements and Declarations}

\medskip
\noindent\textbf{Funding.} This research received no external funding.

\medskip
\noindent\textbf{Competing interests.} The authors declare no competing interests.

\medskip
\noindent\textbf{AI usage.} The authors used Claude Sonnet 4.6 (Anthropic) for assistance with manuscript polishing only. 

\clearpage
\bibliographystyle{plainnat}

\bibliography{cas-refs}

\clearpage
\appendix

\section{Additional Figures}
\label{app:additional_figures}

\subsection{Lambda Calibration Results}

\begin{figure}[h]
  \centering
  \includegraphics[width=\linewidth]{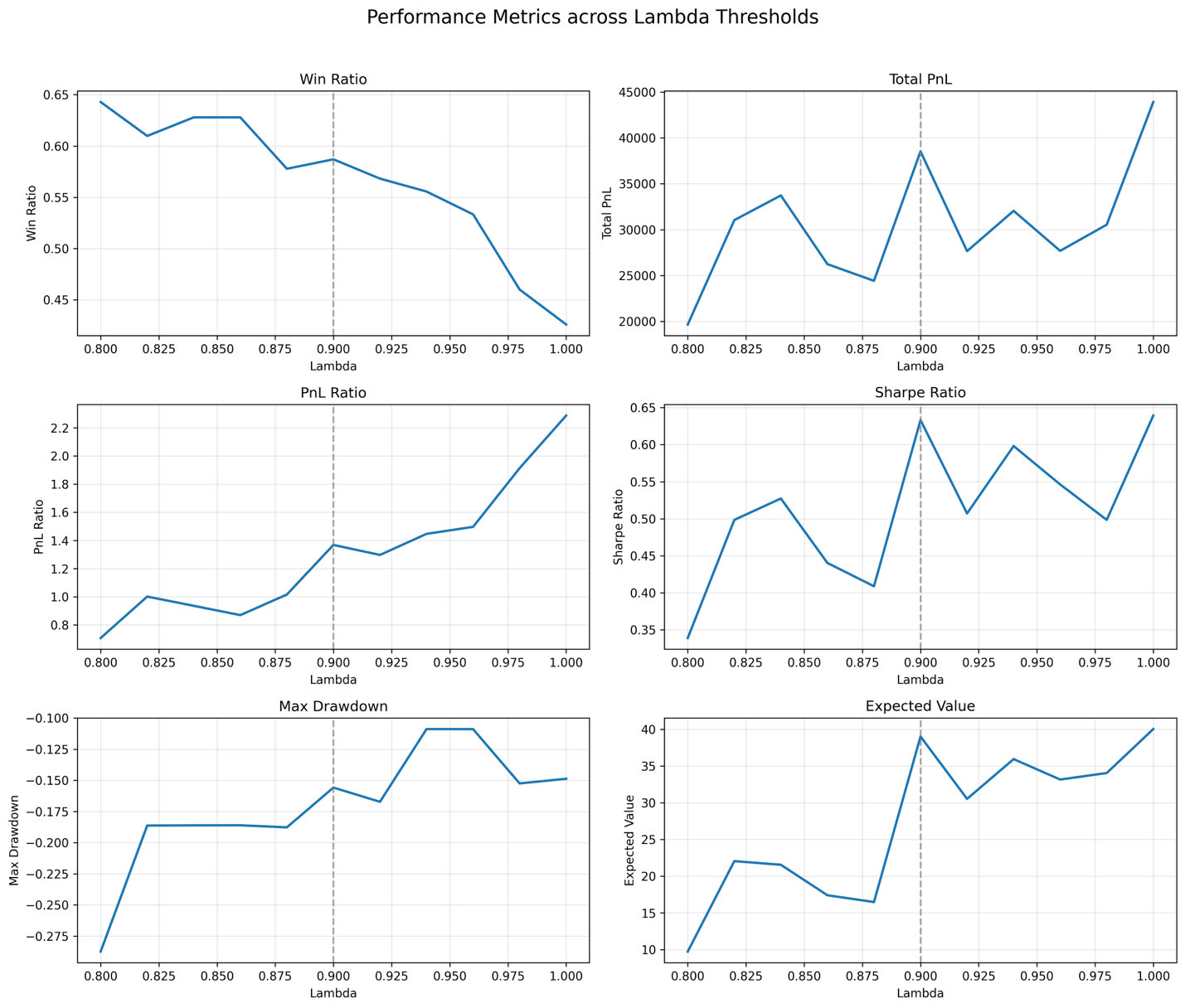}
  \caption{Performance metrics across different $\lambda$ values for MACD strategy for
    SPY.}
  \label{fig:spy}
\end{figure}

\begin{figure}[h]
  \centering
  \includegraphics[width=\linewidth]{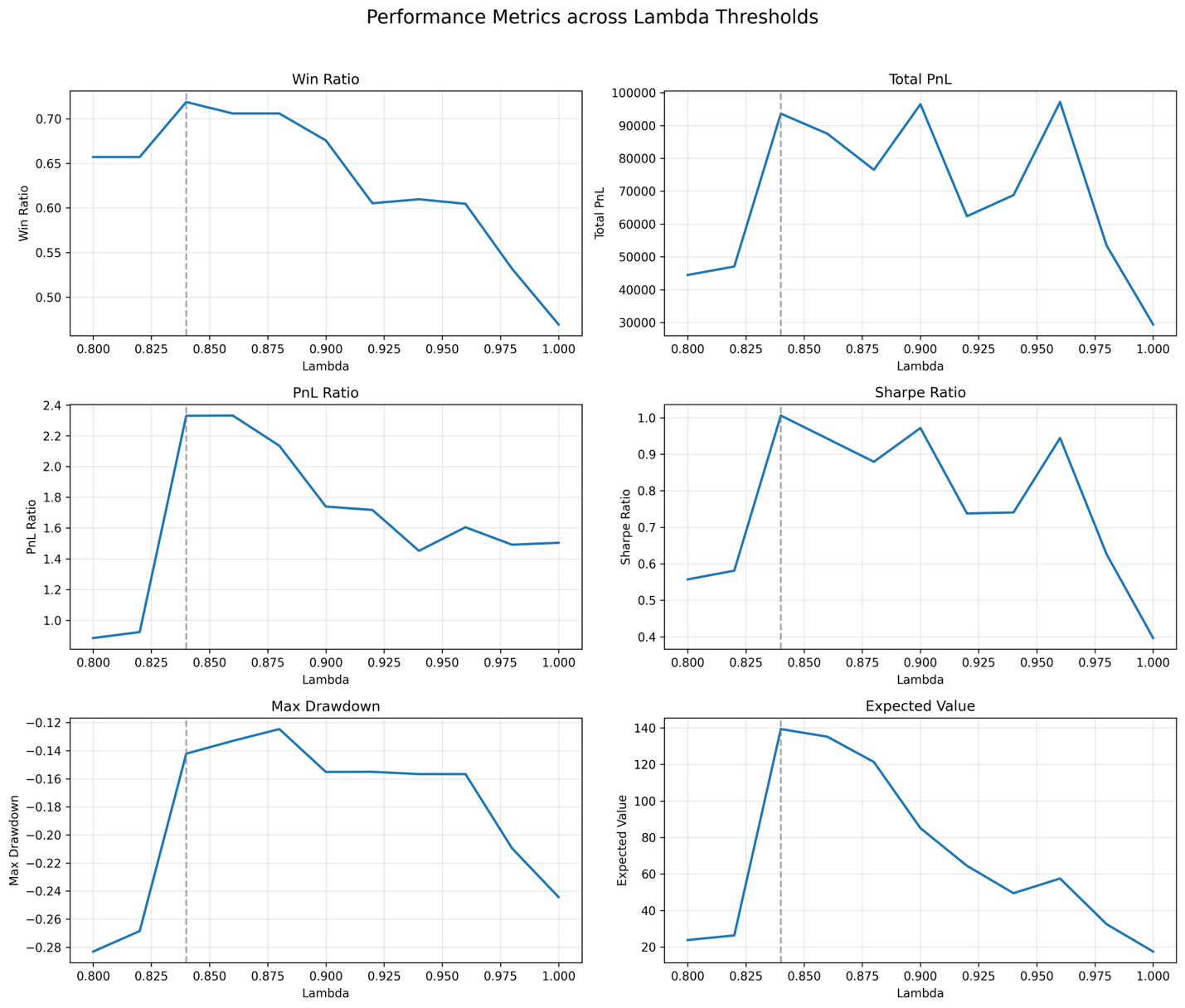}
  \caption{Performance metrics across different $\lambda$ values for MACD strategy for
    QQQ.}
  \label{fig:qqq}
\end{figure}

\begin{figure}[h]
  \centering
  \includegraphics[width=\linewidth]{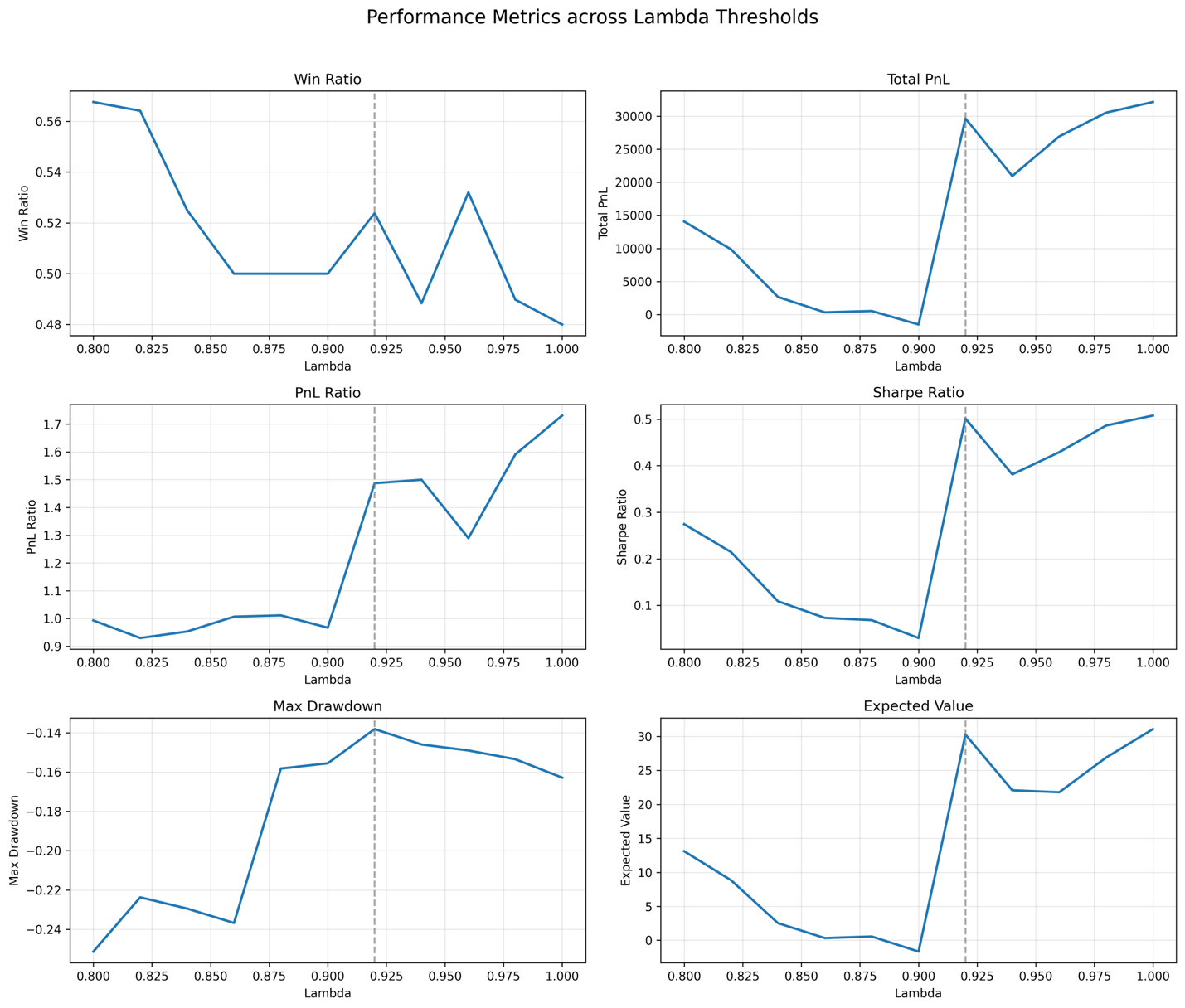}
  \caption{Performance metrics across different $\lambda$ values for MACD strategy for
    DIA.}
  \label{fig:dia}
\end{figure}

\begin{figure}[h]
  \centering
  \includegraphics[width=\linewidth]{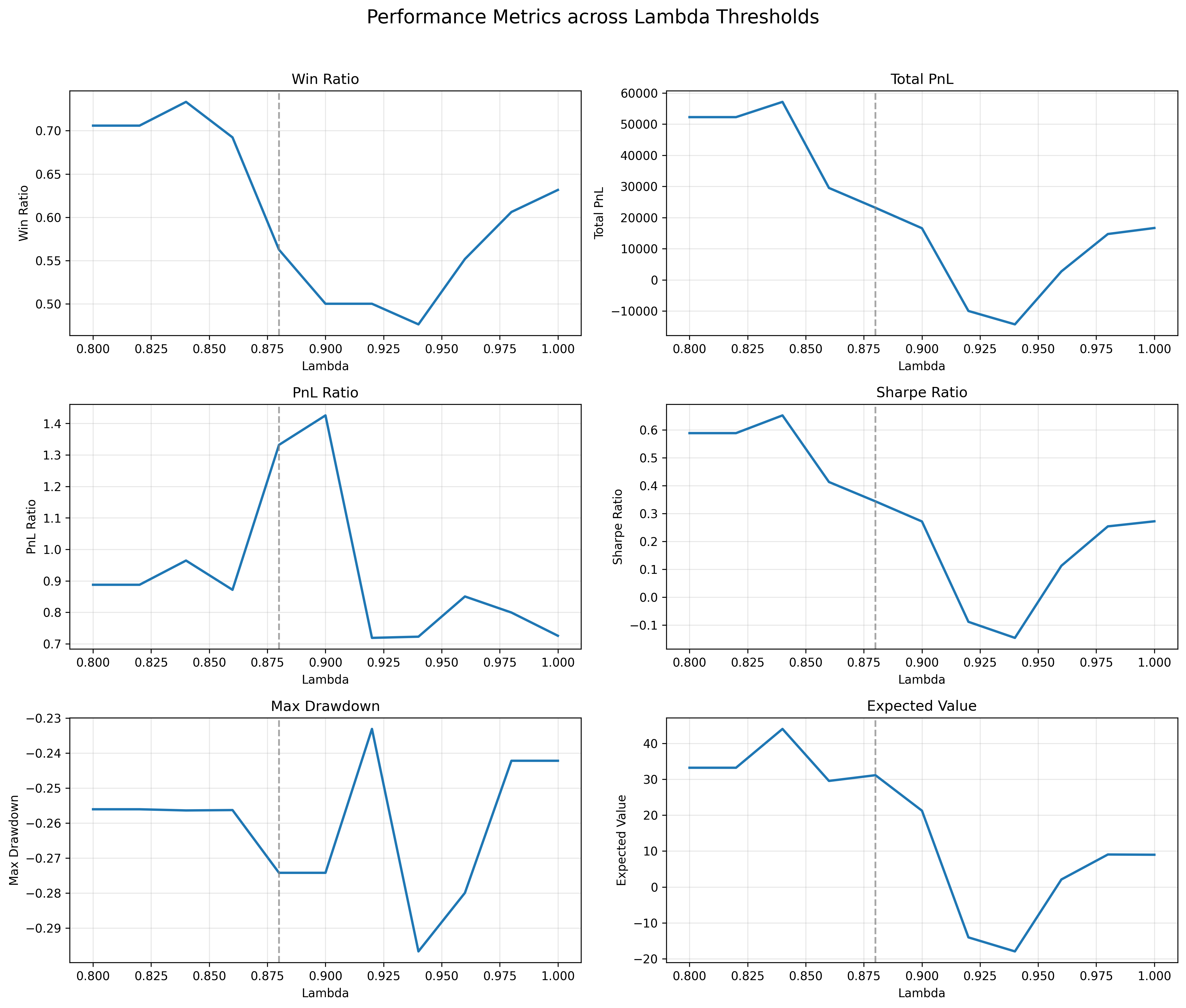}
  \caption{Performance metrics across different $\lambda$ values for the VP-MACD strategy for
    SPY.}
  \label{fig:spy_vp_macd}
\end{figure}

\begin{figure}[h]
  \centering
  \includegraphics[width=\linewidth]{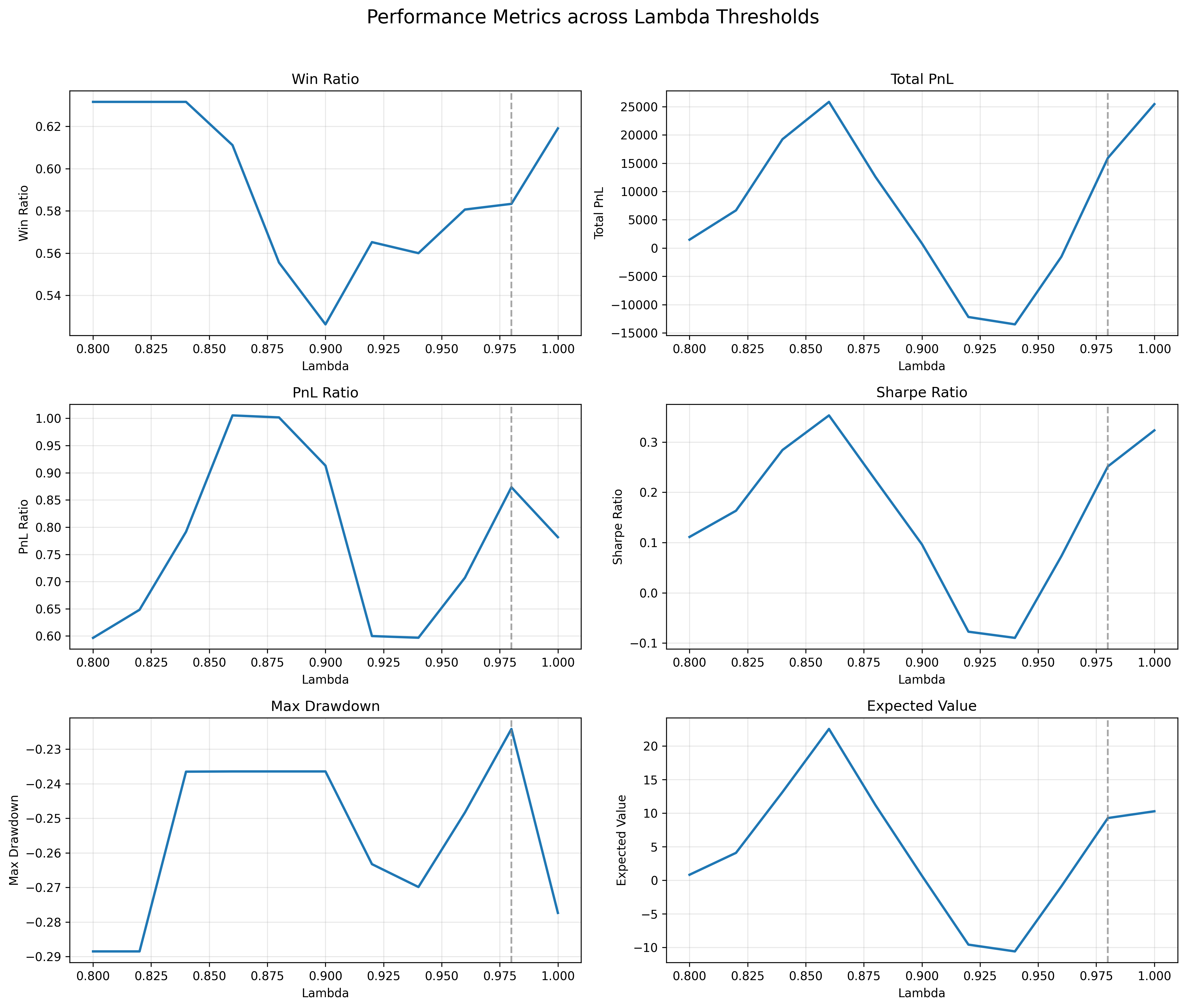}
  \caption{Performance metrics across different $\lambda$ values for the VP-MACD strategy for
    QQQ.}
  \label{fig:qqq_vp_macd}
\end{figure}

\begin{figure}[h]
  \centering
  \includegraphics[width=\linewidth]{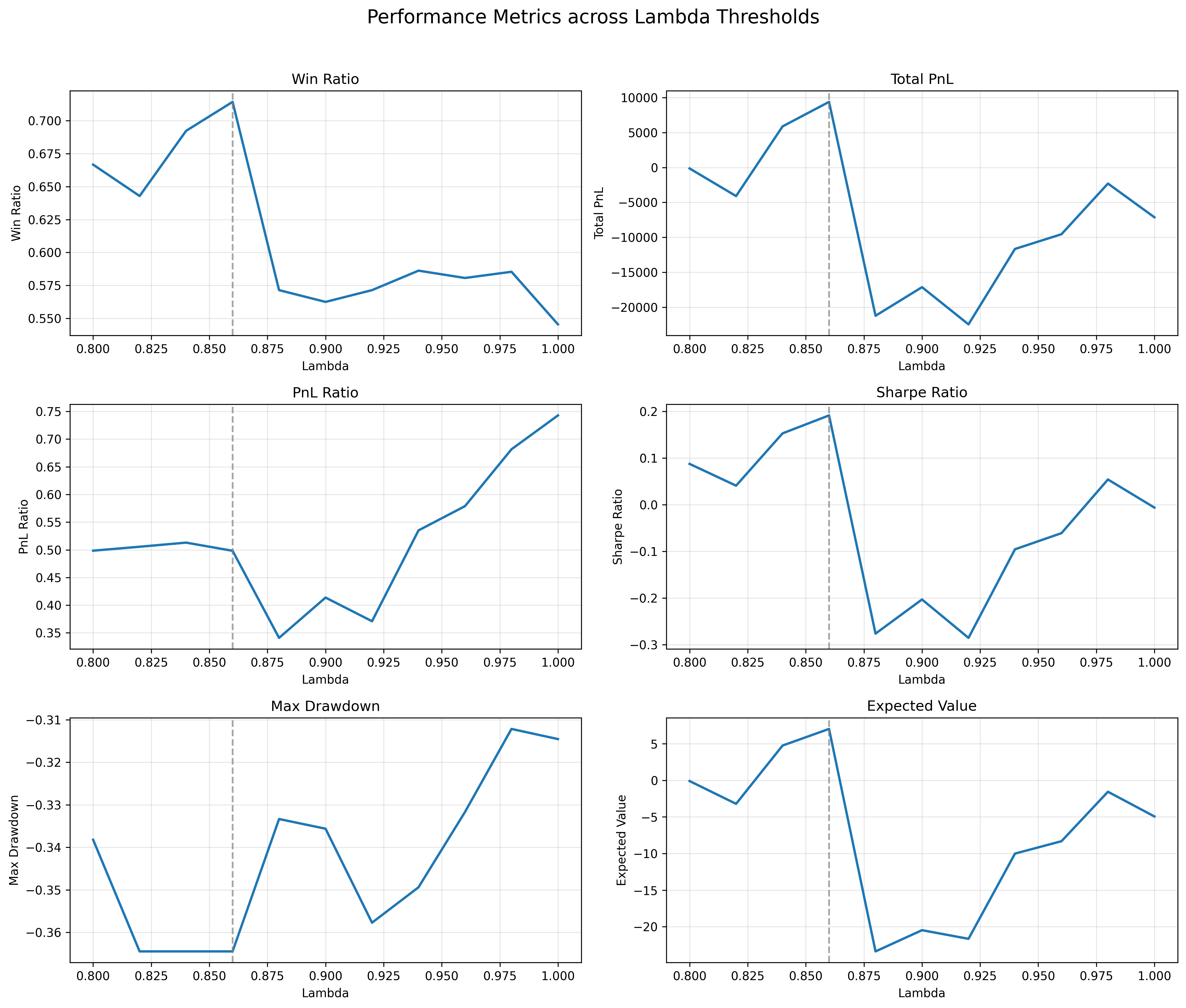}
  \caption{Performance metrics across different $\lambda$ values for the VP-MACD stragegy for
    DIA.}
  \label{fig:dia_vp_macd}
\end{figure}

\FloatBarrier
\subsection{Bootstrap Distributions}

\begin{figure}[h]
  \centering
  \includegraphics[width=\linewidth]{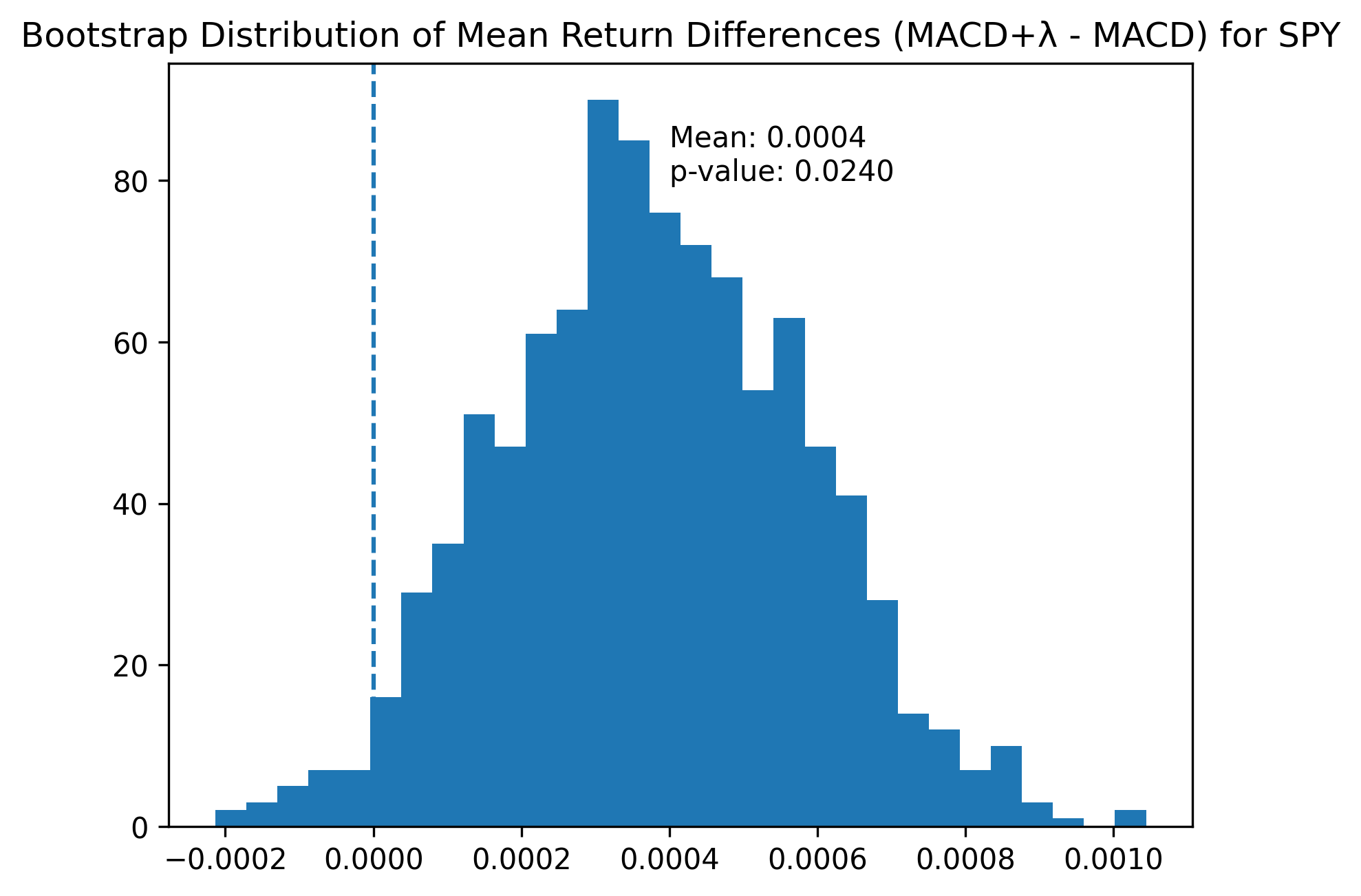}
  \caption{Bootstrap distribution of mean return differences between MACD+$\lambda$ and MACD for SPY.}
  \label{fig:spy_bs}
\end{figure}

\begin{figure}[h]
  \centering
  \includegraphics[width=\linewidth]{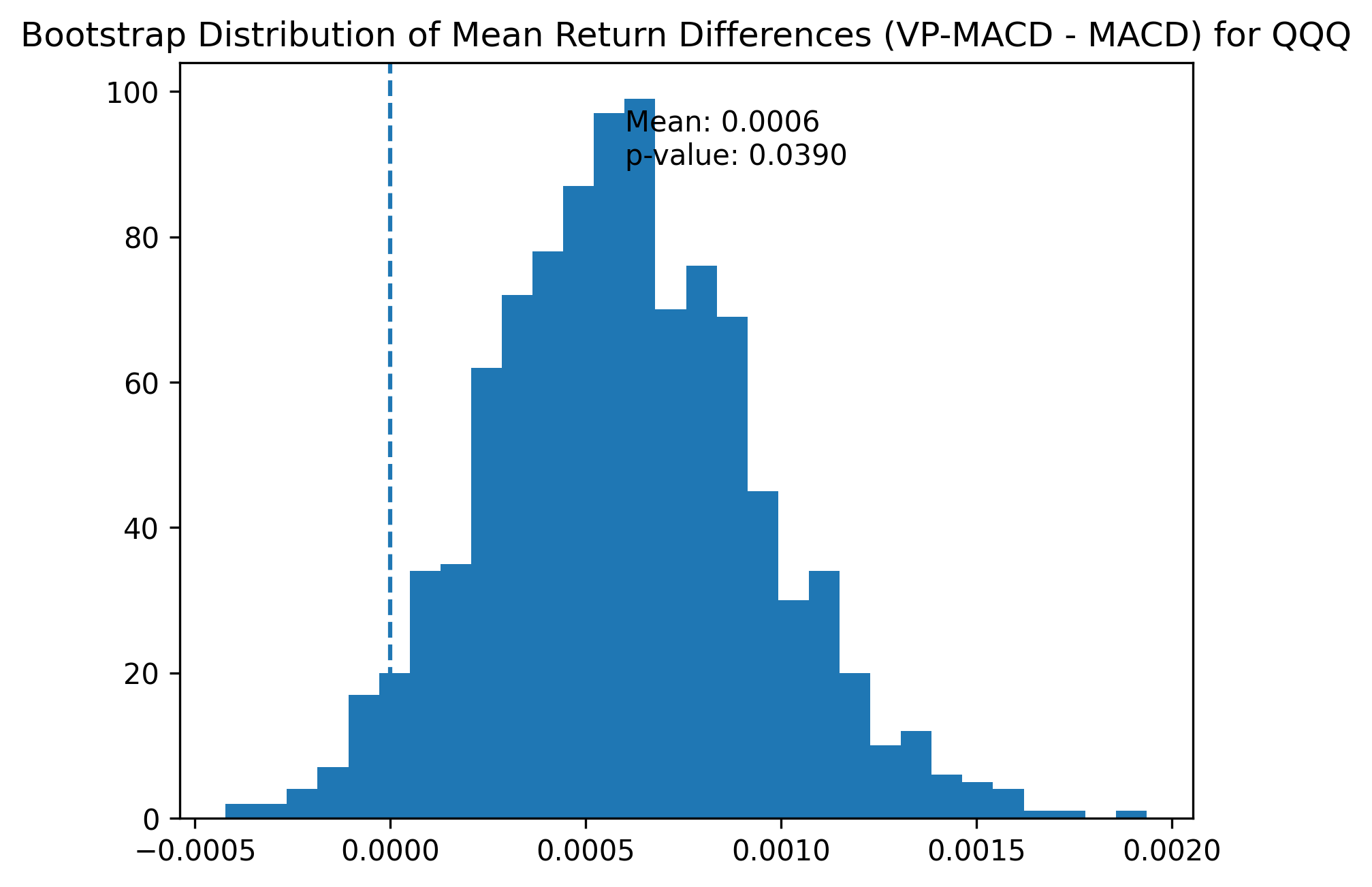}
  \caption{Bootstrap distribution of mean return differences between VP-MACD and MACD for QQQ.}
  \label{fig:qqq_bs}
\end{figure}

\begin{figure}[h]
  \centering
  \includegraphics[width=\linewidth]{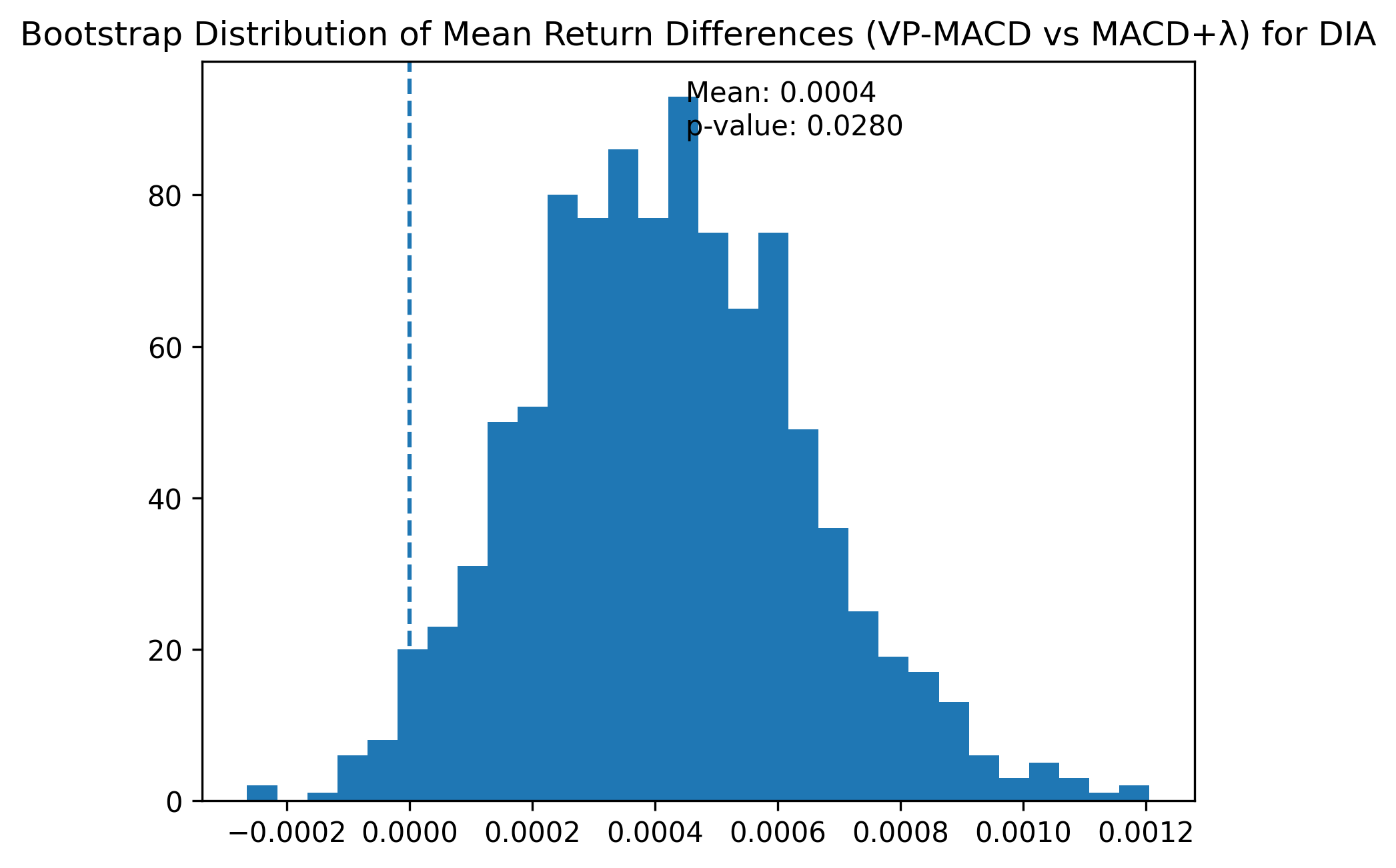}
  \caption{Bootstrap distribution of mean return differences between VP-MACD and MACD+$\lambda$ for DIA.}
  \label{fig:dia_bs}
\end{figure}

\end{document}